\documentclass[twocolumn,pra,showpacs,superscriptaddress]{revtex4}
\usepackage{amssymb}
\usepackage{amsmath}
\usepackage{graphicx}
\usepackage{epstopdf}
\usepackage{subfigure}
\usepackage{natbib}
\usepackage{epsfig}
\usepackage{amsfonts}
\usepackage{mathrsfs}
\usepackage{CJK}
\usepackage[toc,page,title,titletoc,header]{appendix}

\newcommand{\convexhull}{\mathop{\mathrm{co}}}

\def\tr{{\rm tr}}

\begin{document}
\title{Entanglement R\'{e}nyi $\alpha $-entropy}

\author{Yu-Xin Wang}
 \affiliation{School of Physics, Peking University, Beijing 100871, China}

\author{Liang-Zhu Mu}
 \affiliation{School of Physics, Peking University, Beijing 100871, China}

\author{Vlatko Vedral}
\email{vlatko.vedral@gmail.com}
\affiliation{Department of Atomic \& Laser Physics,
Clarendon Laboratory, University of Oxford, Parks Road, Oxford OX1 3PU, UK}
\affiliation{Centre for Quantum Technologies, National University of Singapore, 3 Science Drive 2, Singapore 117543 and\\
Department of Physics, National University of Singapore, 2 Science Drive 3, Singapore 117542}

\author{Heng Fan}
\email{hfan@iphy.ac.cn}
 \affiliation{Institute of Physics,
Chinese Academy of Sciences, Beijing 100190, China}
\affiliation{Collaborative Innovation Center of Quantum Matter, Beijing 100190, China }

\date{\today}
\begin{abstract}
We study the entanglement R\'{e}nyi $\alpha$-entropy (ER$\alpha $E) as the measure of entanglement. Instead of a single quantity in standard entanglement quantification for a quantum state by using the von Neumann entropy for the well-accepted entanglement of formation (EoF),
the ER$\alpha $E gives a continuous spectrum parametrized by variable $\alpha $ as the entanglement measure, and it reduces to the standard EoF in the special case $\alpha \rightarrow 1$. The ER$\alpha $E provides more information in entanglement quantification, and can be used such as in determining the convertibility of entangled states by local operations and classical communication. A series of new results are obtained:
(i) we can show that ER$\alpha $E of two states, which can be mixed or pure, may be incomparable, in contrast to the fact that there always exists an order for EoF of two states; (ii) similar as the case of EoF, we study in a fully analytical way the ER$\alpha $E for arbitrary two-qubit states, the Werner states and isotropic states in general d-dimension; (iii) we provide a proof of the previous conjecture for the analytical functional form of EoF of isotropic states in arbitrary d-dimension.
\end{abstract}

\maketitle

\section{Introduction}
Entanglement is a valuable resource for
quantum information processing \cite{01}. Quantification of
entanglement is a fundamental problem in quantum information science
and quantum physics. Various measures of entanglement have been
proposed such as, entanglement of formation (EoF),
distillable entanglement and entanglement cost \cite{mixedstate1,mixedstate2},
relative entropy of entanglement \cite{VedralPlenio},
see reviews for more results \cite{VedralRMP,PlenioVirmani,HorodeckiE}. These measures have different, yet closely related, physical
interpretations. In general, they can be associated respectively with
different protocols for quantum information processing.
Several well-accepted measures
of entanglement such as EoF  and the relative
entropy of entanglement converge to the same quantity for pure bipartite state, which
is the von Neumann entropy of the reduced density operator of this bipartite state.
For a given state, entanglement measures are not the same in general,
nor a unique quantity even within one kind of measure.
A class of measures may constitute the entanglement monotones with physical
significance in the framework of local operations and classical
communication (LOCC) \cite{NielsenEM,JPLOCC,Vidalmixedqubits,VidalEM}.
R\'{e}nyi $\alpha$-entropy is a natural generalization of von Neumann entropy and
it reduces to the latter when $\alpha $ is approaching 1.

In this paper, we shall consider the quantity entanglement R\'{e}nyi $\alpha$-entropy (ER$\alpha$E) as
the entanglement measure, which is a generalizing of the well-known EoF.
These entropies parameterized by a continuous variable $\alpha $ can be spectrum of entanglement monotone.
Important applications are found by using R\'{e}nyi $\alpha$-entropy in
describing entanglement of ground states of many-body systems
\cite{Turgut,Klimesh,Cui1,Cui2,Cui3,Alioscia1,Alioscia2,Alioscia3},
which are pure states.

In contrast to the relatively simple case of pure entangled states,
the quantification of mixed states entanglement is still challenging due to
the need for hard optimization procedures \cite{HorodeckiE}.
However for EoF, analytical results are well known for
some special cases, including arbitrary two-qubit states based on concurrence in the
seminal work of Wootters \cite{WoottersEoF},
the isotropic states \cite{TerhalEoF} and the Werner states \cite{VWEM} in arbitrary d-dimension.
In parallel with those three analytical results, in this paper, we obtain similarly
analytical results of ER$\alpha $E for those cases. A series of results are obtained,
which are different from and complementary to the EoF in quantifying entanglement.
We show that two-qubit states may be incomparable, which is in contrast with the
expectation by using concurrence that entanglement of two-qubit states can be fully quantified,
which is also different from the case of two-qubit pure states.
We can show that two mixed states may be incomparable. We find that the optimal pure states decompositions
should not be the same, for example, in ER$\alpha $E.
Besides, we provide a proof to the conjecture about the analytical functional form of EoF for isotropic states.

\section{Definition}
Suppose we have a composite system with subsystems
A and B in a pure state $| \psi \rangle $ whose Schmidt decomposition is $| \psi \rangle =\sum_{i = 1}^d \sqrt {\mu _i}|a_i,b_i\rangle _{AB}$ where $\vec{\mu }$ is the Schmidt vector.
For simplicity, we denote the density matrix as $\psi \equiv |\psi \rangle \langle\psi | $,
and let $\rho _{B(A)}= {\rm tr}_{A(B)}(\psi )$ be the reduced density matrix of subsystem B (or A).
The entanglement of pure state $|\psi \rangle $ can be quantified by the
R\'{e}nyi $\alpha $-entropy of one of the reduced density operators,
for example $\rho _{B}$, which is defined as,
\begin{equation}
{\mathcal{R}_\alpha }(\psi ) \equiv {\left({1 - \alpha }\right)}^{-1}\log \left( {{\mathop{\mathrm{tr}}}\rho _B^\alpha } \right),
\label{def}
\end{equation}
As pointed out in the introduction, ${\mathcal{R}_\alpha }(\psi ) $ is reduced to the well known entanglement measure of von Neumann entropy in the $\alpha \to 1$ limit: ${\mathcal{R}_{\alpha \to 1} }(\psi ) =-{\mathrm{tr}}(\rho _B \log \rho _B)$.
This measure of entanglement can be easily generalized to mixed states using the so-called convex roof construction \cite{VidalEM, HorodeckiE}.
For a mixed state with density matrix $\rho$,
 ER$\alpha$E is defined as,
\begin{equation}\label{remixdef}
{\mathcal{R}_\alpha }(\rho ) \equiv \mathop {\min }\limits_{\left\{ {{p_k},{\psi _k}} \right\}} \sum\limits_k {{p_k}{\mathcal{R}_\alpha }\left( {{\psi _k}} \right)}.
\end{equation}
where the minimization is over all possible pure state ensembles $\left\{ {{p_k},{\psi _k}} \right\}$ satisfying ${\rho  = \sum_k {{p_k}{\psi _k}} }$. As in most cases of mixed state entanglement measure, the evaluation of ER$\alpha$E of mixed states is much more difficult to carry out due to the complexity involving
in the optimization.

Our calculation of RE$\alpha$E of symmetric mixed states is closely based on that of two-qubit states, hence at the outset we introduce useful concepts and extend existing results concerning RE$\alpha$E of mixed two-qubit states to a broader range of $\alpha$.

It is well known that
EoF corresponding to $\alpha \rightarrow 1$ for ER$\alpha$E
depends only on concurrence which has an analytic form for arbitrary two-qubit state
and EoF itself can act as a measure of entanglement \cite{WoottersEoF}.
For $\alpha \in \left({1, +\infty}\right)$, ER$\alpha$E
depends similarly only on concurrence \cite{KSRenyi}.
It is then possible that concurrence, in principle,
might be the only essential measure of entanglement for two-qubit state
even ER$\alpha$E for $\alpha \in \left({0,1}\right)$ can act as
entanglement monotones \cite{VidalEM}.
We will show in this paper that this is not the case.
We remark that
ER$\alpha$E satisfies monogamy inequality \cite{CKWmonogamy, OVmonogamy} for multiqubit states when $\alpha =2$ \cite{KSRenyi}.
EoF or concurrence of mixed states
in higher-dimensional system are known for classes of states with special symmetry
such as the Werner states \cite{Wernerstate} and isotropic states \cite{TerhalEoF,RCconcurrence,VWEM}.

\section{The critical value of $\alpha$ for two-qubit state}
Suppose we have a composite system of two-qubit in a pure state $\left| \psi \right \rangle$.
Given the spin flip operation on this state, $| {\tilde \psi }\rangle  = {\sigma _y}| {{\psi ^*}} \rangle$,
its concurrence can be defined as, $\mathcal{C}(\psi )=|\langle {\tilde \psi }|\psi \rangle |$.
For notation simplicity we use $\mathcal{C}$ in the following to denote $\mathcal{C}(\psi )$ or $\mathcal{C}(\psi )$ when no confusion arises.
It is easy to see that the Schmidt coefficients ${\lambda _ \pm }$ of $\left| \psi \right \rangle$, i.e. the eigenvalues of the reduced density matrix $\rho_B$, are in one-to-one correspondence with $\mathcal{C}$ via the relation $\lambda _\pm =(1 \pm \sqrt {1-{\mathcal{C}^2}})/2$.
By direct substitution of ${\lambda _ \pm }$ into the definition Eq.\eqref{def} of
ER$\alpha$E, we obtain
\begin{eqnarray}
{\mathcal{R}_\alpha }(\psi ) &=&(1 - \alpha )^{-1}\log (\lambda _+ ^\alpha +\lambda _-^\alpha )\nonumber\\*
&\equiv & \Omega ({\mathcal{C}},\alpha ),
\label{function}
\end{eqnarray}
where we introduce the function $\Omega ({\mathcal{C}},\alpha )$ for later convenience.

The concurrence of arbitrary mixed state $\rho$ can be similarly defined through the convex roof formula,
$\mathcal{C}(\rho ) \equiv \mathop {\min }_{\left\{ {{p_k},{\psi _k}} \right\}} \sum_k {{p_k}\mathcal{C}\left( {{\psi _k}} \right)}$,
where the minimization is again over all possible pure state decomposition of $\rho$.
An important observation made in \cite{WoottersEoF,HWconcurrence} is that this measure is computable
for two-qubit state. If we generalize the spin flip operation to any mixed state $\rho$ by
$\tilde \rho  = \left( {{\sigma _y} \otimes {\sigma _y}} \right){\rho ^*}\left( {{\sigma _y} \otimes {\sigma _y}} \right)$,
and let ${\Lambda _i^2\left( {{\Lambda _1} \ge {\Lambda _2} \ge {\Lambda _3} \ge {\Lambda _4} \ge 0}\right)}$ denote the eigenvalues of $\tilde \rho \rho $, then the concurrence can be calculated explicitly by
$\mathcal{C}\left( \rho  \right) = \max \left\{ {{\Lambda _1} - {\Lambda _2} - {\Lambda _3} - {\Lambda _4},0} \right\}$.

The ER$\alpha$E of an arbitrary state $\rho$ now becomes
\begin{equation}\label{optimization1}
\mathcal{R}_{\alpha }(\rho ) =\mathop {\min }\limits_{\left\{ {{p_k},{\psi _k}} \right\}} \sum\limits_k {{p_k}\Omega \left( {\mathcal{C}\left( {{\psi _k}} \right),\alpha } \right)}.
\end{equation}
Before we proceed, we first note some existing work on ER$\alpha$E. It is proved in \cite{KSRenyi} that ER$\alpha$E satisfies monogamy inequality \cite{CKWmonogamy, OVmonogamy} for multiqubit states when $\alpha =2$. Also, EoF and concurrence of mixed states
in higher-dimensional system are known for classes of states with special symmetry
such as the Werner states \cite{Wernerstate} and isotropic states \cite{TerhalEoF,RCconcurrence,VWEM}.

Going back to two-qubit system, it is shown in \cite{HWconcurrence,KSRenyi} that for $\alpha \in \left[{1, +\infty}\right)$ this quantity depends only on ${\mathcal{C}}(\rho )$, and in the rest of this section we investigate its behavior in the range of $\alpha \in \left({0,1}\right)$. The most of the calculations, albeit complicating, are non-essential to out understanding of the results. Thus we refer the interested readers to appendix \ref{AppA} and \ref{AppB} for additional details of the calculation, and only discuss the indications of the results we obtain, which form the basis of our calculation of Werner states. We make use of the convexity of the function $\Omega \left( {\mathcal{C},\alpha } \right)$ which is determined from the inequality
\begin{equation}\label{omega}
\frac{{{\partial ^2}\Omega \left( {\mathcal{C},\alpha } \right)}}{{\partial {\mathcal{C}^2}}}
\begin{cases}\le 0,&\alpha  \in \left[ {0,\frac{1}{2}} \right],\\
 \ge 0,& \alpha  \in \left[ {\frac{{\sqrt 7  - 1}}{2}, 1 } \right],\end{cases}
\end{equation}
as well as its monotonicity with respect to $\mathcal{C}$. Thus when $\alpha \ge \alpha _c\equiv \frac{{\sqrt 7  - 1}}{2}\approx 0.82$, the ER$\alpha$E can be calculated analytically
based on concurrence by
\begin{eqnarray}
\mathcal{R}_{\alpha }(\rho )=\Omega ({\mathcal{C}}(\rho ),\alpha ).
\label{maineq}
\end{eqnarray}
The ER$\alpha$E for two-qubit states when $\alpha <\alpha _c$ in general is still a challenging problem.
However, we may instead consider the Werner state which possesses special symmetry \cite{Wernerstate}.

\section{ER$\alpha$E of Werner state}
By the use of the permutation operator  $\mathbb{F} = \sum_{i,j = 1}^d {\left| {ij} \right\rangle \left\langle {ji} \right|}$, the Werner state $\rho _{\scriptscriptstyle{F}}^{\scriptscriptstyle{W}}$ of a bipartite system, consisting of two $d$-dimensional subsystems, can be defined as,
\begin{equation}
\rho _{\scriptscriptstyle{F}}^{\scriptscriptstyle{W}} = \frac{{1 - F}}{2} \frac{{\mathbb{I} + \mathbb{F}}}{{{d^2} + d}} + \frac{{1 + F}}{2}\frac{{\mathbb{I} - \mathbb{F}}}{{{d^2} - d}},
\end{equation}
where the parameter $F \in \left[{-1,1}\right]$ specifying the state can be given from the relation
\begin{equation}
{\mathop{\mathrm{tr}}}\left( {\mathbb{F}\rho _{\scriptscriptstyle{F}}^{\scriptscriptstyle{W}}} \right)= - F.
\end{equation}
Our choice of the parameter $F$ is different from \cite{VWEM} by a minus sign, so that Werner states are separable for $F \le 0$. As we shall show later, ER$\alpha$E with $\alpha = 0$ of the qubit-qubit state is,
interestingly, equal to $F$.

We will make use of the result first obtained in \cite{TerhalEoF}, but we first introduce some notations to simplify the equations. The convex hull of a function $f\left( x \right)$ with $\mathcal{D}$ as its domain is defined as
\begin{equation}
co\left( {f\left( x \right)} \right) 
\equiv \inf  \left\{ 
\sum\limits_k {{p_k}f\left( {x_k} \right) }
\left| {\sum\limits_k {{p_k}{x_k}}  = x,{x_k} \in \mathcal{D}} \right. \right\},
\end{equation}
where the coefficients ${p_k}$ of the convex combinations satisfy $\sum_k {{p_k}}  = 1$; we also need the function ${f_W}\left( \rho  \right) \equiv- \tr\left( {{\mathbb{F}}\rho } \right)$. Thus the RE$\alpha$E of a Werner state is equal to
\begin{equation}\label{reaewco}
{\mathcal{R}_\alpha }\left( {\rho _{\scriptscriptstyle{F}}^{W}} \right) = co\left( {\omega \left( {F,\alpha ,d} \right)} \right),
\end{equation}
where the function $\omega \left( {F,\alpha ,d} \right)$ is defined by $\omega \left( {F,\alpha ,d} \right) = \inf \left\{ {{\mathcal{R}_\alpha }\left( \psi  \right)\left| {{f_{W}}\left( \psi  \right) = F,\, {\rm rk}\left( \psi  \right) \le d} \right.} \right\}$.

To express the value of ${f_{W}}\left( \psi  \right)$, we write the Schmidt decomposition of $\psi$ as
\begin{equation}\label{Schmidtdecom}
\left| \psi  \right\rangle  = \sum_{i = 1}^d {\sqrt {{\mu _i}} \left| {{a_i},{b_i}} \right\rangle }  = \left( {{U_A} \otimes {U_B}} \right)\sum_{i = 1}^d {\sqrt {{\mu _i}} \left| {ii} \right\rangle },
\end{equation}
and let $V \equiv U_A^\dag {U_B},{v_{ij}} \equiv \left\langle i \right|V\left| j \right\rangle $, then we arrive at
\begin{equation}
{f_W}\left( \psi  \right) = -\sum\limits_{i,j = 1}^d {\sqrt {{\mu _i}{\mu _j}} {v_{ji}}v_{ij}^*}.
\end{equation}

We first note that, since the class of Werner state is equivalent to isotropic states for $d=2$, the calculation for that case can alternatively be done for the set of isotropic states, and the results will be identical. The function $\omega\left(F,\alpha \right)$ can be shown to have the explicit formula
\begin{equation}\label{omegaform}
\omega\left(F,\alpha \right)=  \begin{cases}
\Omega\left(F,\alpha \right),& F \in \left({0,1}\right],\\
0, & F \in\left[{-1, 0}\right].
\end{cases}
\end{equation}
Although this is crucial to our computation of RE$\alpha$E of Werner states, the derivation is essentially algebraic and quite tedious. Therefore here we only present the result, and the reader can consult appendix \ref{AppC} for a detailed derivation. Substituting Eq.\eqref{omegaform} into Eq.\eqref{reaewco}, one concludes that
\begin{equation}
{\mathcal{R}_\alpha }\left( {\rho _{\scriptscriptstyle{F}}^{\scriptscriptstyle{W}}} \right)=  \begin{cases}
 co\left(\Omega\left(F,\alpha \right)\right),& F \in \left({0,1}\right],\\
0, & F \in\left[{-1, 0}\right].
\end{cases}
\end{equation}

Thus as far as RE$\alpha$E are concerned, our result implies that the parameter $F$ of Werner states is dimensionless, since Werner states with the same $F$ for given $\alpha$ all have the same value of RE$\alpha$E regardless of the dimension of the Hilbert space. Also, as in the case of isotropic states, we obtain for the class of Werner states a relation between the parameter $F$ and the RE$\alpha$E corresponding to $\alpha=0$
\begin{equation}\label{reaew0}
{\mathcal{R}_{\alpha=0} }\left( {\rho _{\scriptscriptstyle{F}} ^{ \scriptscriptstyle {W} }} \right)=F.
\end{equation}

As an example, we can compare the ER$\alpha $E of a Werner state with $F=0.8$
and a pure state with $F=0.5$, as shown in FIG.1. We note that the pure state here is chosen to be a two-qubit state, in which case if we define $F$ of the said state to be its concurrence, then by Eq.\eqref{function} its RE$\alpha$E coincides in function form with $\omega\left(F,\alpha \right)$, which is shown explicitly in Eq.\eqref{omegaform}. This will simplify the calculation when we perform the compaison, yet the more important reason for using choosing such state is as following: as $d$ increases the value of RE$\alpha$E of a pure state in the $\alpha \to 0$ limit also increases, so the same behavior as in FIG.1 will always be present as long as we set the EoF of that pure state to be low enough without reducing its $d$. Thus we essentially only need to verify the existence of a crossing for the case of Werner states and two-qubit pure states, which is indeed true as demonstrated in FIG. 1: apparently, by the entanglement measure EoF corresponding
to $\alpha =1$, the entanglement of the Werner state is larger than this pure state.
This result seems natural and well-accepted.
Surprisingly, when $\alpha $ is small and is approaching 0,
the order of entanglement for those two states is reversed. One can find that
the entanglement of the pure state is larger than that of the Werner state.

\begin{figure}
\centering
\includegraphics[width=8cm]{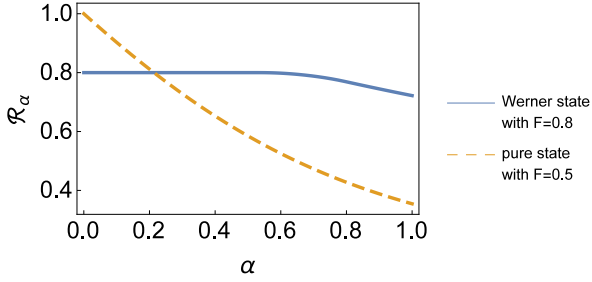}
\caption{
ER$\alpha $E of a Werner state and a pure state.
ER$\alpha $E of the Werner state with $F=0.8$
and ER$\alpha $E of a pure state with $F=0.5$
are presented depending on parameter $\alpha $ as the $x$-axis.
When $\alpha =1$ which is the case of the standard EoF by von Neumann
entropy, the entanglement of Werner state is larger than that of the pure state.
However, the order of the entanglement is reversed when $\alpha $ is close to zero. The specific choice of and definition of $F$ for the pure state are discussed in the main text.}
\label{wernerpure}
\end{figure}

Actually, from Eq.\eqref{reaew0} we have that the entanglement of Werner states will always be less than that of an
arbitrary pure entangled state in the limit of $\alpha \rightarrow 0$.
Following the discussions above, we only need to show that for the case of $d=2$. The optimal pure states decomposition for a Werner state will include a maximally entangled state with probability $F$ and the identity operator, resulting in that ER$\alpha $E equals to $F$, while ER$\alpha $E of a generic entangled pure states is 1.

In this sense, we find that ER$\alpha $E of Werner states and pure states may be incomparable.

\section{ER$\alpha $E of isotropic state}
The class of isotropic states, specified by a parameter $F \in \left[{0,1}\right]$, consists of convex mixtures of a maximally entangled state and a maximally mixed state
\begin{equation}
\rho _{\scriptscriptstyle{F}} ^{iso} = F{P_ + } + \frac{{1 - F}}{{{d^2} - 1}}\left( {\mathbb{I} - {P_ + }} \right),\quad F \in \left[{0,1}\right],
\end{equation}
where ${P_ + }$ is the projector onto the subspace spanned by the maximally entangled state $\left| {{\Psi ^ + }} \right\rangle =\frac{{1 }}{{\sqrt d }} \sum_{i = 1}^d {\left| {ii} \right\rangle} $. We can define an isotropic state analog of ${f_W}\left( \rho  \right)$ as ${f_{{\Psi ^ + }}}\left( \rho  \right) =\left\langle {{\Psi ^ + }} \right|\rho \left| {{\Psi ^ + }} \right\rangle  =\tr \left( {{P_ + }\rho } \right)$, which is the fidelity between $\rho$ and ${\Psi ^ +}$.

Again let us start with the formula
\begin{equation}
{\mathcal{R}_\alpha }\left( {\rho _{\scriptscriptstyle{F}}^{iso}} \right) = co\left( {\eta \left( {F,\alpha ,d} \right)} \right),
\end{equation}
where the function $\eta \left( {F,\alpha ,d} \right)$ is defined as $\eta \left( {F,\alpha ,d} \right) = \inf \left\{ {{\mathcal{R}_\alpha }\left( \psi  \right)\left| {{f_{{\Psi ^ + }}}\left( \psi  \right) = F,\, {\rm rk}\left( \psi  \right) \le d} \right.} \right\}$. Making use of the Schmidt decomposition in Eq.\eqref{Schmidtdecom}, and define $W \equiv U_A^{\rm T} {U_B},{w_{ij}} \equiv \left\langle i \right|W\left| j \right\rangle $, then straightforward calculation yields
\begin{equation}
{f_{{\Psi ^ + }}}\left( {{{\psi}} } \right) = \frac{1}{d}{\left| {\sum\limits_{i = 1}^d {\sqrt {{\mu _i}} {w_{ii}}} } \right|^2}.
\end{equation}

The value of $\eta \left( {F,\alpha ,d} \right)$ for $F \in \left[ {0,\frac{1}{d}} \right]$ can then be easily deduced by setting ${\mu _1} = 1,{w_{11}} = \sqrt F $, which yields $\eta\left( {F,\alpha ,d} \right) =0$. For $F \in \left( {\frac{1}{d},1} \right]$, using the method of Lagrange multipliers, we derive a closed expression of the function ${\eta\left( {F,\alpha ,d} \right)}$ as
\begin{equation}\label{etac}
\eta\left( {F,\alpha ,d} \right) = \frac{1}{{1 - \alpha }}\log \left[ {{\gamma ^\alpha } + {{\left( {d - 1} \right)}^{1 - \alpha }}{{\left( {1 - \gamma } \right)}^\alpha }} \right],
\end{equation}
where $\gamma$, here standing for the function $\gamma \left( {F,d} \right)$, is defined as $\gamma \left( {F,d} \right) \equiv \frac{{1}}{d} {{\left( {\sqrt F  + \sqrt {\left( {d - 1} \right)\left( {1 - F} \right)} } \right)}^2}$. In the limit of $\alpha \to 1$, $\eta\left( {F,\alpha ,d} \right)$ reduces to the function
\begin{equation}\label{epsilon}
\varepsilon \left( {F,d} \right) = {H_2}\left( {\gamma } \right) + \left( {1 - \gamma } \right)\log \left( {d - 1} \right),
\end{equation}
where $H_2\left( \cdot \right)$ denotes the binary entropy function. This result, which we have derived here as a special case, is first obtained in \cite{TerhalEoF}; since the calculation of $\eta\left( {F,\alpha ,d} \right)$ is just a straightforward generalization, we do not go into the calculation details of Eq. \eqref{etac}. As an application of the results we obtain here, we provide in the appendix \ref{AppD} an analytical proof of the conjecture of EoF of isotropic states in \cite{TerhalEoF}, namely
\begin{equation}
\mathcal{E}\left( {\rho _{\scriptscriptstyle{F}}^{iso}} \right)=\label{T1}
\begin{cases} 0,&F \in \left[ {0,\frac{1}{d}} \right],\\
\varepsilon \left( {F,d} \right),&F \in \left[ {\frac{1}{d},\frac{{4\left( {d - 1} \right)}}{{{d^2}}}} \right],\\
\frac{{d\log \left( {d - 1} \right)}}{{d - 2}}\left( {F - 1} \right) + \log d,&F \in \left[ {\frac{{4\left( {d - 1} \right)}}{{{d^2}}},1} \right].\end{cases}
\end{equation}

\begin{figure}[h]
\centering
\includegraphics[width=8cm]{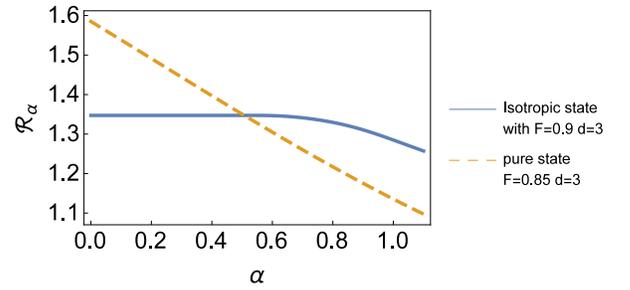}
\caption{(color online) Comparing the RE$\alpha$E with $\alpha \in \left[{0,1}\right]$ of an isotropic state and a pure state.}
\label{Fig2}
\end{figure}

Concluding this section we present an example shown in FIG.\ref{Fig2} obtained by numerical evaluation,
of a comparison between the R\'{e}nyi entropy of a certain entangled isotropic state of $F=0.85$ and a pure state chosen such that its EoF is smaller than that of the former. We further note that the crossing behavior is also possible between two mixed states, although it does not necessarily have to be always present, as indicated in FIG.3: there's evidently no crossing between the two isotropic states with $d=3$, yet the two with $F=0.9$, $d=2$ and $F=0.7$ and $d=3$ respectively have crossed RE$\alpha$E.

\begin{figure}[h]
\centering
\includegraphics[width=8cm]{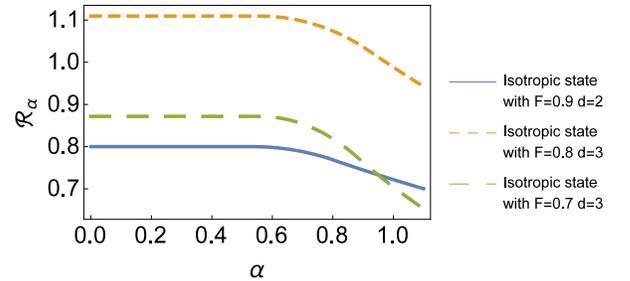}
\caption{(color online) Comparing the RE$\alpha$E with $\alpha \in \left[{0,1}\right]$ of mixed states.}
\label{Fig3}
\end{figure}

We also note that our result provides a nice corollary, namely the relation between the parameter $F =\tr \left( {{P_ + }\rho } \right)$ specifying the class of isotropic states and its Renyi entropy with $\alpha=0$
\begin{equation}
{\mathcal{R}_{\alpha=0} }\left( {\rho _{\scriptscriptstyle{F}} ^{ iso }} \right)=\frac{{Fd-1}}{{d-1}} \log d.
\end{equation}

\section{Discussions}
ER$\alpha $E quantifies entanglement. For two-qubit states, when $\alpha \ge \alpha _c$,
ER$\alpha $E can be obtained analytically based on the well-known concurrence.
This result implies that the pure states in the optimal decomposition for ER$\alpha $E when $\alpha \ge \alpha _c$
possess the same Schmidt vector similar as that for concurrence.
The general analytical formula of ER$\alpha $E even for the simplest two-qubit states
is still a challenging problem, also the pure states decomposition in general will not possess
the same Schmidt vector.
However, ER$\alpha $E for Werner states can be obtained.
Interestingly, we notice that ER$\alpha $E for the simplest two-qubit states
may be incomparable, implying that they are not local convertible by LOCC. This phenomenon is previously
only known for higher-dimensional systems. Analytical results of ER$\alpha $E for Werner states and isotropic
states are obtained. A series of new phenomena are found by using ER$\alpha $E,
which may stimulate more interests in studying quantum entanglement quantification.

{\it{Acknowledgements.}}---This work was supported the NSFC under
grant No. 91536108, the grant from Chinese Academy of Sciences (XDB01010000).
 V. V. acknowledges funding from
the EPSRC, the Templeton Foundation, the Leverhulme
Trust, the Oxford Martin School, the National Research
Foundation (Singapore), the Ministry of Education (Singapore)
and from the EU Collaborative Project TherMiQ
(Grant Agreement 618074). V.V. is also partially supported by
Chinese Academy of Sciences President's International Fellowship Initiative.

\appendix

\section{Proof of the general framework of computing ER${\alpha}$E of states with local symmetry}
\label{AppA}
A general approach is derived for systematically calculating entanglement monotone of states with local symmetry in \cite{VWEM}. Here we present the main steps in the calculations, and refer the interested reader to the paper mentioned above for its application in calculating entanglement measures other than ER${\alpha}$E, while the latter is the topic of the main text. Suppose we have a set of states $\rho^{\mathcal{G}}$ invariant under a group,  $\mathcal{G}$, of local unitaries, which by definition satisfies the condition
\begin{equation}\left[ {{\rho ^{\mathcal{G}}},U \otimes V} \right] = 0,\quad \forall \left( {U \otimes V} \right) \in {\mathcal{G}}.\end{equation}
The twirling operation ${\mathcal{P}^{\mathcal{G}}}$ is a well-established tool \cite{TerhalEoF}, which is defined as
\begin{equation}
{\mathcal{P}^{\mathcal{G}}}\rho  = \int\limits_{\left( {U \otimes V} \right) \in {\mathcal{G}}} {dU\left( {U \otimes V} \right)\rho {{\left( {U \otimes V} \right)}^\dag }}.
\end{equation}
Noting that twirling can be thought of as a probabilistic superposition of unitaries, it thus belongs to the class of local operation and classical communication (LOCC), and by definition an entanglement monotone of a generic state decreases or remains the same under such an operation.

By change of the integration variable, it is easy to show that $\left[ {{\mathcal{P}^{\mathcal{G}}}\rho ,U \otimes V} \right] = 0$ for all $\left( {U \otimes V} \right) \in {\mathcal{G}}$, which means that ${\mathcal{P}^{\mathcal{G}}}\rho$ also exhibits the symmetry described by group $\mathcal{G}$. Therefore, the result of twirling ${\mathcal{P}^{\mathcal{G}}}$ depends solely on the fidelities between the twirled state and a certain set of projectors, which are associated with the group of symmetry operations  $\mathcal{G}$. This set of quantities (described by a vector in \cite{VWEM}) can also be used as a unique specifier of the states within the class having corresponding symmetry. For isotropic and Werner states, the set of projectors can be simplified to only one operator. Since these are the only sets of states we consider in this paper, we can restrict our discussion to this special case, where we define a function ${f_{\mathcal{G}}}\left( \rho  \right)$ with the property that
\begin{equation}
{\mathcal{P}^{\mathcal{G}}}\rho  = \rho _{{F = {f_{\mathcal{G}}}\left( \rho  \right)}}^{\mathcal{G}},
\end{equation}
and the function ${f_{\mathcal{G}}}\left( \rho  \right)$ is a one-to-one mapping from the class of symmetrical states to the set of real numbers (or in general, real vectors).

Now we move on to the calculation of an entanglement monotone: suppose $\mathcal{X}\left( \psi  \right)$ is defined on the set of pure states, where $\psi$ stand for the corresponding density operator, and we use convex roof construction \cite{HorodeckiE} to generalize this measure to mixed state $\rho$ by
\begin{equation}\label{convexrodef}
\mathcal{X}\left( \rho  \right) \equiv \inf \left\{ {\sum\limits_k {{p_k}\mathcal{X}\left( {{\psi _k}} \right)} \left| {\sum\limits_k {{p_k}{\psi _k}}  = \rho } \right.} \right\}.
\end{equation}
Then by defintion of the infimum, we have
\begin{align}{\mathcal{X}}\left( {\rho _F^{\mathcal{G}}} \right) =& \inf \left\{ {\sum\limits_k {{p_k}\mathcal{X}\left( {{\psi _k}} \right)} \left| {\sum\limits_k {{p_k}{\psi _k}}  = \rho _F^{\mathcal{G}}} \right.} \right\}, \nonumber\\
\ge& \inf \left\{ {\sum\limits_k {{p_k}\mathcal{X}\left( {{\psi _k}} \right)} \left| {{\mathcal{P}^{\mathcal{G}}}\left( {\sum\limits_k {{p_k}{\psi _k}} } \right) = \rho _F^{\mathcal{G}}} \right.} \right\}, \nonumber\\
\Rightarrow{\mathcal{X}}\left( {\rho _F^{\mathcal{G}}} \right)& \ge \inf \left\{ {\sum\limits_k {{p_k}\mathcal{X}\left( {{\psi _k}} \right)} \left| {\sum\limits_k {{p_k}{f_{\mathcal{G}}}\left( {{\psi _k}} \right)}   = F} \right.} \right\}.\label{ineq1}\end{align}
Let $\left\{ {p_k^{\left( {\rm{0}} \right)},  {\psi _k^{\left( {\rm{0}} \right)}} } \right\}$ be a pure state ensemble which is the optimal decomposition achieving the infimum in the right hand side, namely
\begin{align}
&\sum\limits_k {p_k^{\left( {\rm{0}} \right)}{f_{\mathcal{G}}}\left( {\psi _k^{\left( {\rm{0}} \right)}} \right)}  = F,\\
\sum\limits_k {p_k^{\left( {\rm{0}} \right)}\mathcal{X}\left( {\psi _k^{\left( {\rm{0}} \right)}} \right)}  &= \inf \left\{ {\sum\limits_k {{p_k}\mathcal{X}\left( {{\psi _k}} \right)} \left| {\sum\limits_k {{p_k}{f_{\mathcal{G}}}\left( {{\psi _k}} \right)}  = F} \right.} \right\}.\label{sumeqinf}
\end{align}
It follows that the twirling of the new state $\rho ' = \sum\limits_k {p_k^{\left( {\rm{0}} \right)}\psi _k^{\left( {\rm{0}} \right)}} $ is just $\rho _F^{\mathcal{G}}$,
\begin{align}
&{\mathcal{P}^{\mathcal{G}}}\rho ' = {\mathcal{P}^{\mathcal{G}}}\left( {\sum\limits_k {p_k^{\left( {\rm{0}} \right)}\psi _k^{\left( {\rm{0}} \right)}} } \right) \nonumber\\
&= \rho _{F = {f_{\mathcal{G}}}\left( {\rho '} \right)}^{\mathcal{G}} = \rho _{F = \sum\limits_k {p_k^{\left( {\rm{0}} \right)}{f_{\mathcal{G}}}\left( {\psi _k^{\left( {\rm{0}} \right)}} \right)} }^{\mathcal{G}} = \rho _F^{\mathcal{G}},
\end{align}
and from the fact that twirling opration is a LOCC operation, we obtain the inequality
\begin{equation}\label{ineq2}
\mathcal{X}\left( {\rho '} \right) \ge \mathcal{X}\left( {{\mathcal{P}^{\mathcal{G}}}\rho '} \right) = \mathcal{X}\left( {\rho _F^{\mathcal{G}}} \right).
\end{equation}

On the other hand, by the property of the convex roof construction in Eq.\eqref{convexrodef}, we have
\begin{equation}
\sum\limits_k {p_k^{\left( {\rm{0}} \right)}\mathcal{X}\left( {\psi _k^{\left( {\rm{0}} \right)}} \right)}  \ge \mathcal{X}\left( {\sum\limits_k {p_k^{\left( {\rm{0}} \right)}\psi _k^{\left( {\rm{0}} \right)}} } \right) = \mathcal{X}\left( {\rho '} \right),
\end{equation}
which combined with Eqs.\eqref{ineq1}, \eqref{sumeqinf} and \eqref{ineq2} provides a tight bound for $\mathcal{X}\left( {\rho _F^{\mathcal{G}}} \right)$ for explicitly calculation
\begin{align}
&\mathcal{X}\left( {\rho _F^{\mathcal{G}}} \right)=\mathcal{X}\left( {\rho '} \right)=\sum\limits_k {p_k^{\left( {\rm{0}} \right)}\mathcal{X}\left( {\psi _k^{\left( {\rm{0}} \right)}} \right)}\nonumber \\
=& \inf \left\{ {\sum\limits_k {{p_k}\mathcal{X}\left( {{\psi _k}} \right)} \left| {\sum\limits_k {{p_k}{f_{\mathcal{G}}}\left( {{\psi _k}} \right)}  = F} \right.} \right\}.
\end{align}

Now we've derived a directly calculable formula of $\mathcal{X}\left( {\rho _F^{\mathcal{G}}} \right)$, which can be computed via a two-step procedure. The first step involves an optimization over pure states with the same value of the function ${f_{\mathcal{G}}}\left( \rho  \right)$,
\begin{equation}
\chi \left( F \right) = \inf \left\{ {\mathcal{X}\left( \psi  \right)\left| {{f_{\mathcal{G}}}\left( \psi  \right) = F} \right.} \right\}.
\end{equation}
While this function is evaluted in \cite{VWEM} for entanglement of formation of both Werner and isotropic states, we note here that it is not immediately clear how it should be computed for any particular class of states or type of entanglement monotone. The calculation still needs to be done explicitly for specific combinations, given the general framework.

The next step is theoretically more easy and computable in general, but in practice does not always yield elegant or explicit results without numerical calculation. We define the convex hull of a function $f\left(x\right)$ defined on $\mathcal{D}$ to be
\begin{align}
&co\left( {f\left( x \right)} \right) =\nonumber\\
& \inf \left\{ {\sum\limits_k {{p_k}f\left( {{x_k}} \right)} \left| {\sum\limits_k {{p_k}{x_k}}  = x,\sum\limits_k {{p_k}}  = 1,{x_k} \in \mathcal{D}} \right.} \right\},
\end{align}
and the entanglement monotone $\mathcal{X}\left( {\rho _F^{\mathcal{G}}} \right)$ of state ${\rho _F^{\mathcal{G}}}$ can be shown to be
\begin{equation}
\mathcal{X}\left( {\rho _F^{\mathcal{G}}} \right) = co\left( {\chi \left( F \right)} \right).
\end{equation}
It is easy to see that this step is computable in general.

\section{Proof of convexity property of the function $\Omega \left( {\mathcal{C},\alpha } \right)$}
\label{AppB}
The function $\Omega \left( {\mathcal{C},\alpha } \right)$ is defined as
\begin{equation}
\Omega ({\mathcal{C}},\alpha )\equiv(1 - \alpha )^{-1}\log (\lambda _+ ^\alpha +\lambda _-^\alpha )  ,
\label{functionapp}
\end{equation}
where $\lambda _\pm =(1 \pm \sqrt {1-{\mathcal{C}^2}})/2$.
Let us present the first derivative of $\Omega \left( {\mathcal{C},\alpha } \right)$ with respect to $\mathcal{C}$, ${{\partial \Omega }}/{{\partial \mathcal{C}}}$, in terms of the Schmidt coefficients ${\lambda _ \pm }$ by using Eq.(\ref{functionapp}),
\begin{equation}\label{pOpC}
\frac{{\partial \Omega (\mathcal{C},\alpha )}}{{\partial \mathcal{C}}}
= \frac{\alpha }{(1 - \alpha )(\lambda _+^\alpha +\lambda _-^\alpha )}
[\lambda _+^{\alpha -1}\frac{{\rm d}\lambda _+ }{{\rm d}\mathcal{C}}+
\lambda _-^{\alpha -1}\frac{{\rm d}\lambda _- }{{\rm d}\mathcal{C}}].
\end{equation}
The derivatives of ${\lambda _ \pm }$ are,
\begin{subequations}\label{derivativesimp}
\begin{eqnarray}
\frac{{\rm d}{\lambda _\pm }}{{\rm d}\mathcal{C}}&= &\frac{\mathcal{C}}{{2\left( {1 - 2{\lambda _ \pm }} \right)}},\\
\frac{{{\rm d}^2}{\lambda _\pm}}{{\rm d}{\mathcal{C}^2}} & =& \frac{2}
{{\mathcal{C}^2}(1 - 2\lambda _\pm )}
\frac{{\rm d}{\lambda _\pm }}{{\rm d}\mathcal{C}}.
\end{eqnarray}
\end{subequations}
Also we know,
${\rm d}{\lambda _ +}/{\rm d}{\mathcal{C}}+ {\rm d}\lambda _-/{\rm d}\mathcal{C} = 0$,
and we introduce the notations, $D_1 = \left|{\rm d}\lambda _+/{\rm d}\mathcal{C}\right| =
{\mathcal{C}}/{2\sqrt {1 -{\mathcal{C}^2}} }$ and $x \equiv \lambda _-/\lambda _+$.
Substitute Eq.\eqref{derivativesimp} into the derivative in Eq.\eqref{pOpC}, we arrive at
\begin{equation}
\frac{{\partial \Omega \left( {\mathcal{C},\alpha } \right)}}{{\partial \mathcal{C}}}  = \frac{{\alpha \lambda _ + ^{\alpha  - 1}{D_1}}}{{\left( {\alpha  - 1} \right)\left( {\lambda _ + ^\alpha  + \lambda _ - ^\alpha } \right)}}\left( {1 - {x^{\alpha  - 1}}} \right) \ge 0.
\end{equation}
Namely $\Omega ({\mathcal{C},\alpha })$ is a monotonically increasing function with respect to $\mathcal{C}$.

The evaluation of the second derivative of $\Omega ({\mathcal{C},\alpha })$
is more complicated,
\begin{eqnarray}
\frac{{{\partial ^2}\Omega \left( {\mathcal{C},\alpha } \right)}}{{\partial {\mathcal{C}^2}}}
= -\frac{{\alpha {\lambda _ + ^{2\alpha  - 2} } {D_1^2}}}{{\left( { 1 - \alpha} \right){{\left( {\lambda _ + ^\alpha  + \lambda _ - ^\alpha } \right)}^2}}}K,
\end{eqnarray}
where
\begin{eqnarray}
K&=&{{{\left( {1 - {x^{\alpha  - 1}}} \right)}^2} +\frac{{{{\left( {1 + x} \right)}^2}}}{{2x\left( {1 - x} \right)}}g\left( {x,\alpha } \right)},\label{defiK}
\\
 g\left( {x,\alpha } \right) &=&1 - {x^{2\alpha  - 1}} - \left( {2\alpha  - 1} \right)\left( {1 - x} \right){x^{\alpha  - 1}},\label{defig}
\end{eqnarray}
and $x \in \left( {0,1} \right] $ and $\alpha \in \left[ {0,1} \right)$.
Observing that
\begin{eqnarray}
\begin{cases}K\ge  {\left( {1 - {x^{\alpha  - 1}}} \right)^2} \ge  0,& {\alpha \in \left[ {0,\frac{1}{2}} \right],}\\
K\le   {\left( {1 - {x^{\alpha  - 1}}} \right)^2,} &  {\alpha \in\left( {\frac{1}{2},1} \right)},\end{cases}
\end{eqnarray}
we thus conclude that the function $\Omega \left( {\mathcal{C},\alpha } \right)$  is concave with respect to $\mathcal{C}$ for $\alpha \in \left[ {0,\frac{1}{2}} \right]$,
\begin{equation}
\frac{{{\partial ^2} \Omega\left( {\mathcal{C},\alpha } \right)}}{{\partial {\mathcal{C}^2}}} \le -\frac{{\alpha \lambda _ + ^{2\alpha  - 2} D_1^2 {\left( {1 - {x^{\alpha  - 1}}} \right)^2}}}{{\left( {1 - \alpha } \right){{\left( {\lambda _ + ^\alpha  + \lambda _ - ^\alpha }\right)}^2}}} \le 0.
\end{equation}
This result is actually opposite of the convexity of the function
$\Omega \left( {\mathcal{C},\alpha } \right)$ for $\alpha \in \left({1, +\infty}\right)$.
We thus give a negative answer for the holding of the relation
\begin{eqnarray}
\mathcal{R}_{\alpha }(\rho )=\Omega ({\mathcal{C}}(\rho ),\alpha ),
\label{maineqapp}
\end{eqnarray}
for $\alpha \in [0,\frac{1}{2}]$.

Next, we consider the region $\alpha \in\left( {\frac{1}{2},1} \right)$.
For a fixed $\alpha $, the second derivative of $\Omega \left( {\mathcal{C},\alpha } \right)$
may have a zero corresponding to $\mathcal{C}_0$ in the interval $\mathcal{C}_0 \in\left[ {0,1} \right]$.
Numerical calculation shows that the value of $\mathcal{C}_0$ increases monotonically with respect to $\alpha$,
see FIG. 1 for the dependence of $\mathcal{C}_0$ on $\alpha $.
So there may exist a critical value of $\alpha$ corresponding to $\mathcal{C} =1$
such that the second derivative of $\Omega \left( {\mathcal{C},\alpha } \right)$ is zero.
Such a critical value $\alpha_c$ does exist, such that the
simplification, meaning the holding of Eq.(\ref{maineq}),
is still valid for any $\alpha$ larger than this value.
In fact, it is not difficult to obtain the value of $\alpha_c$ analytically.
One simply considers the limit $\mathcal{C} \to 1$ and the requirement that,
\begin{equation}
\mathop {\lim }\limits_{\mathcal{C} \to 1} \frac{{{\partial ^2}\Omega \left( {\mathcal{C},\alpha } \right)}}{{\partial {\mathcal{C}^2}}}  \ge 0,
\end{equation}
which is equivalent to $\mathop {\lim }\limits_{x \to 1} K\le 0$.
Referring to the definitions of $K$ and $g(x,\alpha )$ in Eqs.\eqref{defiK} and \eqref{defig}, we derive the following inequality,
\begin{equation}
\frac{{\left( {\alpha  - 1} \right)}}{3}\left[ {3\left( {\alpha  - 1} \right) + \left( {2\alpha  - 1} \right)\alpha } \right] \le 0.
\end{equation}
The value of $\alpha_c$ can be calculated by considering the condition for equality in the above expression,
which gives us,
\begin{equation}
\alpha_c=\frac{{\sqrt 7  - 1}}{2}\approx 0.82.
\end{equation}
This solution is consistent with our numerical result presented in FIG. 1.

\begin{figure}
\centering
\includegraphics[width=7cm]{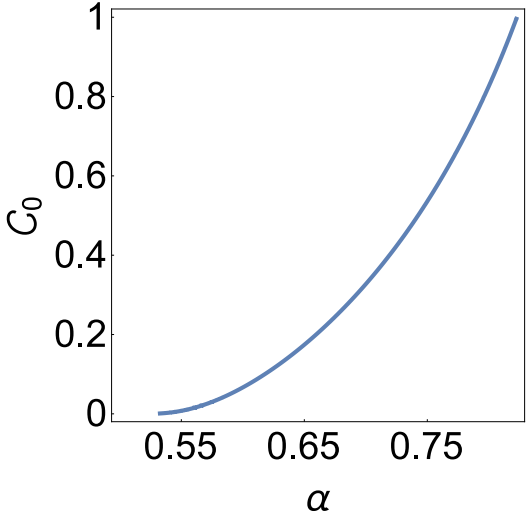}
\caption{(color online) By using condition
$\frac{{{\partial ^2}\Omega \left( {\mathcal{C},\alpha } \right)}}{{\partial {\mathcal{C}^2}}} =0$,
we can find the dependence of $\mathcal{C}_0$, which satisfies this equation, on $\alpha $.}
\label{critical}
\end{figure}

\section{Derivation of the closed form of the function $\omega\left( {F,\alpha ,d} \right)$}
\label{AppC}
Here we give a detailed proof of the formula
\begin{equation}
\omega\left( {F,\alpha ,d} \right) =  \begin{cases}
\Omega\left(F,\alpha \right), &F \ge 0,\\0, &F < 0.
\end{cases}
\end{equation}

The value of $\omega \left( {F,\alpha ,d} \right)$ for $F \in\left[{-1, 0}\right]$ can be easily obtained by setting ${\mu _1} = 1,{v_{11}} = \sqrt F $ in the explicit expression of the function
\begin{equation}\label{fWofpsi}
{f_W}\left( \psi  \right) = -\sum\limits_{i,j = 1}^d {\sqrt {{\mu _i}{\mu _j}} {v_{ji}}v_{ij}^*} ,
\end{equation}
giving $\omega \left( {F,\alpha ,d} \right)=0$, which reproduces the separability condition of Werner state.
While the optimization in the definition of $\omega \left( {F,\alpha ,d} \right) = \inf \left\{ {{\mathcal{R}_\alpha }\left( \psi  \right)\left| {{f_W}\left( \psi  \right) = F,\, {\rm rk}\left( \psi  \right) \le d} \right.} \right\}$ for general $d$ cannot be carried out in a straightforward fashion, it is rather simple for two-qubit state.
Setting $d=2$ in Eq.\eqref{fWofpsi}, we have that for $F>0$,
\begin{equation}
\mathcal{C}\left(\psi\right)=2\sqrt {{\mu _1}{\mu _2}}\ge F,
\end{equation}
where equality holds if and only if ${\mu _{1,2}}=\lambda_\pm$. Since the function $\Omega\left(F,\alpha \right)$ is monotonically increasing with respect to $F$, we thus obtain the inequality $\Omega\left(\mathcal{C}\left(\psi\right),\alpha \right)={\mathcal{R}_\alpha }\left( \psi \right)\ge\Omega\left(F,\alpha \right)$. It then follows that,
\begin{equation}
\omega\left( {F,\alpha ,d} \right) =  \begin{cases}
\Omega\left(F,\alpha \right), &F \ge 0,\\0, &F < 0.
\end{cases}
\end{equation}

When seen as a real vector, $\vec{\mu}=\left[{\mu_i}\right]$ belongs to the convex set of real, at most $d$-dimensional vectors such that $\sum_{i = 1}^d {{\mu _i} = 1({\mu _i} \ge 0})$, which we denote by ${\mathcal{K}_d}$. It follows that ${f_W}\left( \psi  \right)$ can also be seen as a function ${f_W}\left( \vec{\mu}, V \right)$ of a real vector $\vec{\mu}$ and a unitary matrix $V$. For $F \in \left({0,1}\right] $, we can calculate the quantity defined by
\begin{equation}\label{taudef}
\tau \left( {F,\alpha ,d} \right)\equiv \inf \left\{{\sum\limits_i {\mu _i^\alpha } } \left| {{f_W}\left( \vec{\mu}, V \right)  =  - F, \vec{\mu} \in {\mathcal{K}_d} } \right. \right\},
\end{equation}
from which the function $\omega \left( {F,\alpha ,d} \right)$ can be obtained from $\omega \left( {F,\alpha ,d} \right)=\log \left[\tau \left( {F,\alpha ,d} \right)\right]$, by the monotonicity of the logarithm function. Noting that the minimization in Eq. \eqref{taudef}, with $d$ larger than 2, always covers all possible combinations of $\left\{ {{\mu _i},{v_{ij}}} \right\}$ with $d=2$, we have
\begin{equation} \label{bound2}
\tau \left( {F,\alpha ,d} \right) \le \tau \left( {F,\alpha ,2} \right).
\end{equation}
Physically this implies that the amount of entanglement contained in Werner states is upper bounded by that of a maximally entangled pair of qubits (for example the Bell states). A consequence of this constraint is that when equality in the infimum in Eq. \eqref{taudef} is achieved, the largest ${\mu_i}$, which we denote by ${\mu _{\max }}$, must be bigger or equal to $\frac{1}{2}$. In fact, if ${\mu _{\max }}$ is smaller than $\frac{1}{2}$, we will derive a contradiction of Eq. \eqref{bound2}
\begin{equation}{\mu _{\max }} < \frac{{\rm{1}}}{\rm{2}} \Rightarrow\tau \left( {F,\alpha ,2} \right)\le \tau \left( {1,\alpha ,2} \right)< {\sum\limits_i {\mu _i^\alpha } }.\end{equation}
Thus we have a constraint on ${\mu _{\max }}$ that reads
\begin{equation}\label{boundmu}{\mu _{\max }} \ge \frac{1}{2}.\end{equation}

Our next step is to seek an upper bound of ${\mu _{\max }}$, which combined with Eq. \eqref{boundmu} will further constrain the possible values that
$\omega \left( {F,\alpha ,d} \right)$ may take. First we derive some inequalities that the set of variables $\left\{ {{\mu _i},{v_{ij}}} \right\}$ must satisfy. Noting that $ \sum_{i = 1}^d {{\mu _i}} =1$, we have
\begin{align}
1 &= \sum\limits_{i = 1}^d {{\mu _i}}  = \sum\limits_{i,j} {{\mu _i}{{\left| {{v_{ji}}} \right|}^2}} \nonumber \\
= & {\mu _{{i_m}}} + \sum\limits_{j \ne {i_m}} {{\mu _j}{{\left| {{v_{{i_m}j}}} \right|}^2}}  + \sum\limits_{i,j \ne {i_m}} {{\mu _i}{{\left| {{v_{ji}}} \right|}^2}}=1 ,
\end{align}
where ${i_m}$ is one of the indices such that ${\mu _{{i_m}}}={\mu _{max}}$. With the use of Cauchy-Schwarz inequality, it follows that
\begin{widetext}
\begin{align}&{\left| {\sum\limits_{\mathop {i \ne j}\limits_{i,j \ne {i_m}} }^d {\sqrt {{\mu _i}{\mu _j}} {v_{ji}}v_{ij}^*} } \right|^2} \le \left( {\sum\limits_{\mathop {i \ne j}\limits_{i,j \ne {i_m}} }^d {{{\left| {\sqrt {{\mu _i}} {v_{ji}}} \right|}^2}} } \right)\left( {\sum\limits_{\mathop {i \ne j}\limits_{i,j \ne {i_m}} }^d {{{\left| {\sqrt {{\mu _j}} v_{ij}^*} \right|}^2}} } \right) = {\left( {\sum\limits_{\mathop {i \ne j}\limits_{i,j \ne {i_m}} }^d {{\mu _i}{{\left| {{v_{ji}}} \right|}^2}} } \right)^2},\\
&{\left| {\sum\limits_{j \ne {i_m}}^d {\sqrt {{\mu _{{i_m}}}{\mu _j}} {v_{j{i_m}}}v_{{i_m}j}^*} } \right|^2} \le {\left( {\sum\limits_{j \ne {i_m}}^d {{\mu _{{i_m}}}{{\left| {{v_{j{i_m}}}} \right|}^2}} } \right)\left( {\sum\limits_{j \ne {i_m}}^d {{\mu _j}{{\left| {v_{{i_m}j}^*} \right|}^2}} } \right)} .  \end{align}\end{widetext}
Thus the following inequalities can be easily derived
\begin{align}
&{\left| {\sum\limits_{\mathop {i \ne j}\limits_{i,j \ne {i_m}} } {\sqrt {{\mu _i}{\mu _j}} {v_{ji}}v_{ij}^*} } \right|}\le  {\sum\limits_{\mathop {i \ne j}\limits_{i,j \ne {i_m}} } {{\mu _i}{{\left| {{v_{ji}}} \right|}^2}} },\label{inefirst}\\
&\left| {\sum\limits_{j \ne {i_m}} {\sqrt {{\mu _{{i_m}}}{\mu _j}} {v_{j{i_m}}}v_{{i_m}j}^*} } \right| \le \sqrt {st}\label{inesecond}  ,
\end{align}
where for later convenience we set
\begin{align}
s&= {\sum\limits_{j \ne {i_m}} {{\mu _{{i_m}}}{{\left| {{v_{j{i_m}}}} \right|}^2}} } ,\label{values}\\
t&= {\sum\limits_{j \ne {i_m}} {{\mu _j}{{\left| {v_{{i_m}j}^*} \right|}^2}} }\label{valuet} .
\end{align}

Making use of the condition in the infimum Eq. \eqref{taudef}, namely ${f_W}\left( \vec{\mu}, V \right)   = \sum_{i = 1}^d {{\mu _i}{{\left| {{v_{ii}}} \right|}^2}}  + \sum_{i \ne j} {\sqrt {{\mu _i}{\mu _j}} {v_{ji}}v_{ij}^*}  =  - F$, we have for $F \in \left[ {0,1} \right]$ that
\begin{align}
& F + \sum\limits_{i = 1}^d {{\mu _i}{{\left| {{v_{ii}}} \right|}^2}} = \left| {\sum\limits_{i \ne j}^d {\sqrt {{\mu _i}{\mu _j}} {v_{ji}}v_{ij}^*} } \right| \\
\le& 2\left| {\sum\limits_{j \ne {i_m}}^d {\sqrt {{\mu _{{i_m}}}{\mu _j}} {v_{j{i_m}}}v_{{i_m}j}^*} } \right| + \left| {\sum\limits_{\mathop {i \ne j}\limits_{i,j \ne {i_m}} }^d {\sqrt {{\mu _i}{\mu _j}} {v_{ji}}v_{ij}^*} } \right|\nonumber
\end{align}\begin{align}
 \le &2\left| {\sum\limits_{j \ne {i_m}}^d {\sqrt {{\mu _{{i_m}}}{\mu _j}} {v_{j{i_m}}}v_{{i_m}j}^*} } \right| + \sum\limits_{\mathop {i \ne j}\limits_{i,j \ne {i_m}} }^d {{\mu _i}{{\left| {{v_{ji}}} \right|}^2}}\nonumber\\
=& 1 - {\mu _{{i_m}}} - \sum\limits_{j \ne {i_m}} {{\mu _j}{{\left| {{v_{{i_m}j}}} \right|}^2}}  + 2\left| {\sum\limits_{j \ne {i_m}}^d {\sqrt {{\mu _{{i_m}}}{\mu _j}} {v_{j{i_m}}}v_{{i_m}j}^*} } \right|\nonumber\\
\le &1 - {\mu _{{i_m}}} - \left( {\sum\limits_{j \ne {i_m}}^d {{\mu _j}{{\left| {v_{{i_m}j}^*} \right|}^2}} } \right) + 2\sqrt {st}\nonumber\label{quadraticoft}\\
 = &1 - {\mu _{{i_m}}} - t + 2\sqrt {t\left( {\sum\limits_{j \ne {i_m}}^d {{\mu _{{i_m}}}{{\left| {{v_{j{i_m}}}} \right|}^2}} } \right)}\end{align}

To further evaluate the expression in Eq.\eqref{quadraticoft}, we note that this is essentially a quadratic function, with the argument $\sqrt{t}$ satisfying the inequality
\begin{equation}t = \sum\limits_{j \ne {i_m}} {{\mu _j}{{\left| {v_{{i_m}j}^*} \right|}^2}}  \le \sum\limits_{j \ne {i_m}} {{\mu _j}}  = 1 - {\mu _{{i_m}}},\end{equation}
and we consider two possible cases of $\sqrt{t}$ to derive a uniform bound. First, if
\begin{equation}
{{\sum\limits_{j \ne {i_m}} {{\mu _{{i_m}}}{{\left| {{v_{j{i_m}}}} \right|}^2}} }=\mu _{{i_m}}} - {\mu _{{i_m}}}{\left| {{v_{{i_m}{i_m}}}} \right|^2} \ge 1 - {\mu _{{i_m}}} ,
\end{equation}
then we have $\sum_{j \ne {i_m}} {{\mu _j}{{\left| {v_{{i_m}j}^*} \right|}^2} }  \le 1 - {\mu _{{i_m}}} \le {\mu _{{i_m}}} - {\mu _{{i_m}}}{\left| {{v_{{i_m}{i_m}}}} \right|^2}$ such that
\begin{equation}
 F + \sum\limits_{i = 1}^d {{\mu _i}{{\left| {{v_{ii}}} \right|}^2}}  \le 2\sqrt {{\mu _{\max }}\left( {1 - {\mu _{\max }}} \right)}.
\end{equation}
If instead the variables $\left\{ {{\mu _i},{v_{ij}}} \right\}$ satisfy
\begin{equation}
{\sum\limits_{j \ne {i_m}}^d {{\mu _{{i_m}}}{{\left| {{v_{j{i_m}}}} \right|}^2}} }={\mu _{{i_m}}} - {\mu _{{i_m}}}{\left| {{v_{{i_m}{i_m}}}} \right|^2} < 1 - {\mu _{{i_m}}},
\end{equation}
then we obtain, similarly,
\begin{align}
 &1 - {\mu _{{i_m}}}{\left| {{v_{{i_m}{i_m}}}} \right|^2} < 2\left( {1 - {\mu _{{i_m}}}} \right),\\
\Rightarrow& \sum\limits_{j \ne {i_m}}^d {{\mu _j}{{\left| {v_{{i_m}j}^*} \right|}^2} \le {\mu _{{i_m}}} - {\mu _{{i_m}}}{{\left| {{v_{{i_m}{i_m}}}} \right|}^2}}  < 1 - {\mu _{{i_m}}},\\
\Rightarrow& F + \sum\limits_{i = 1}^d {{\mu _i}{{\left| {{v_{ii}}} \right|}^2}}  \le 1 - {\mu _{{i_m}}}{\left| {{v_{{i_m}{i_m}}}} \right|^2} < 2\left( {1 - {\mu _{{i_m}}}} \right), \\
\Rightarrow& F + \sum\limits_{i = 1}^d {{\mu _i}{{\left| {{v_{ii}}} \right|}^2}}  \le 2\sqrt {{\mu _{\max }}\left( {1 - {\mu _{\max }}} \right)}.
\end{align}

Combining the two cases, and noting that ${\mu _{{i_m}}}={\mu _{max}}$, we derive a constraint of $\mu_{\max}$ in terms of $F$
\begin{equation}\label{T2}
F \le F + \sum\limits_{i = 1}^d {{\mu _i}{{\left| {{v_{ii}}} \right|}^2}}  \le 2\sqrt {{\mu _{\max }}\left( {1 - {\mu _{\max }}} \right)},
\end{equation}
Let $\mu \left( {F} \right)={\frac{{1 + \sqrt {1 - {F^2}} }}{2}}$ (which for simplicity we will denote by $\mu$ when no confusion arises), and Eqs. \eqref{boundmu} and \eqref{T2} give ${\mu _{\max }} \le \mu \left( {F} \right)$. Making use of the Schur concavity of the function $\tau \left( {F,\alpha ,d} \right)$, we have
\begin{equation}\tau \left( {F,\alpha ,d} \right) = \sum\limits_i {\mu _i^\alpha }  \ge \mu _{\max }^\alpha  + {\left( {1 - {\mu _{\max }}} \right)^\alpha } \ge  \tau \left( {F,\alpha ,2} \right). \end{equation}
On the other hand we have already obtained the inequality \eqref{bound2}, and together the two bounds yield
\begin{equation}
\tau \left( {F,\alpha ,d} \right) = \sum\limits_i {\mu _i^\alpha }  \ge  \tau \left( {F,\alpha ,2} \right) \ge\tau \left( {F,\alpha ,d} \right), \end{equation}
indicating that the bound given in Eq.\eqref{bound2} is indeed tight. We have now succeeded in calculating the functions $\tau \left( {F,\alpha ,d} \right) $ and $\omega\left( {F,\alpha ,d} \right)$
\begin{align}
\tau \left( {F,\alpha ,d} \right) &= \tau \left( {F,\alpha ,2} \right)\\
\omega\left( {F,\alpha ,d} \right) &=  \begin{cases}
\Omega\left(F,\alpha \right)&F \ge 0\\0 &F < 0\end{cases}
\end{align}

\section{Proof of the conjecture on EoF of isotropic state}
\label{AppD}
We first state the rigorous result on entanglement of formation of isotropic state ${\rho _{\scriptscriptstyle{F}}^{iso}}$ in $d$ dimensional Hilbert space, which we denote by $\mathcal{E}\left( {\rho _{\scriptscriptstyle{F}}^{iso}} \right)$
\begin{equation}\label{chexp}
\mathcal{E}\left( {\rho _F^{iso}} \right) = \convexhull\left( {\varepsilon \left( {F,d} \right)} \right)
\end{equation}
where $\convexhull \left( \cdot  \right)$ denotes the convex hull of a function, and the function $\varepsilon \left( {F,d} \right)$ is found to be
\begin{align}
\varepsilon\left( {F,d} \right) &= {H_2}\left( {\gamma \left( {F,d} \right)} \right) + \left( {1 - \gamma \left( {F,d} \right)} \right)\log \left( {d - 1} \right)\\
\gamma \left( {F,d} \right) &= \frac{1}{d} {{{\left( {\sqrt F  + \sqrt {\left( {d - 1} \right)\left( {1 - F} \right)} } \right)}^2}} \label{gamma0}
\end{align}
and ${H_2}\left( x \right)$ is the binary entropy function.

In Vollbrecht and Terhal's work \cite{TerhalEoF}, a closed expression of EoF of isotropic states is conjectured,
whose validity the authors argue can always be verified for any given $d$, by directly plotting the function $\varepsilon \left( {F,d} \right)$ and examining its behavior. The rigorous proof of $d=3$ case is provided.
It is now generally accepted that the conjecture is true for arbitrary $d$,
it seems that a proof is still necessary.
Alternatively, here we seek to prove this conjecture in an analytical fashion, without any presumption about the value of $d$.

The proof consists of two steps: we first prove a general statement about $\varepsilon \left( {F,d} \right)$, when treating $F$ as the argument and $d$ as a parameter, and the next step simply involves the verification of a criterion of the point $F=\frac{{4\left( {d - 1} \right)}}{{{d^2}}}$. First we wish to show that, the concavity of $\varepsilon \left( {F,d} \right)$ with respect to $F$ on the interval $F \in \left[{1,0}\right]$ is in general analogous to that of the special case with $d=3$, i.e. first concave upwards and then concave downwards. We will prove this statement by directly evaluating the value of second derivative of $\varepsilon \left( {F,d} \right)$ with respect to $F$. To keep the equations as simple as possible, we use $\varepsilon$ and $ \gamma$ to denote respectively the functions
\begin{align}
\varepsilon \left( {F,d} \right) &= {H_2}\left( {\gamma } \right) + \left( {1 - \gamma } \right)\log \left( {d - 1} \right),\\
\gamma \left( {F,d} \right) &= \frac{1}{d} {{{\left[ {\sqrt F  + \sqrt {\left( {d - 1} \right)\left( {1 - F} \right)} } \right]}^2}} .\label{gamma}
\end{align}
A bit of algebra gives us
\begin{align}
\frac{{d\gamma }}{{dF}} &=  - \frac{{\sqrt {\gamma \left( {1 - \gamma } \right)} }}{{\sqrt {F\left( {1 - F} \right)} }},\\
\frac{{{d^2}\gamma }}{{d{F^2}}}  &= - \frac{{\sqrt {d - 1} }}{{2d\gamma \left( {1 - \gamma } \right)\sqrt {F\left( {1 - F} \right)} }} {{\left( {\frac{{d\gamma }}{{dF}}} \right)}^2}.
\end{align}
Substituting the derivatives of $\gamma$ with respect to $F$ into the relation
\begin{equation}
\frac{{{\partial ^2}\varepsilon}}{{\partial {F^2}}} = \frac{{{\partial ^2 \varepsilon}}}{{\partial {\gamma ^2}}} {\left( {\frac{{d\gamma }}{{dF}}} \right)}^2 + \frac{{{d^2}\gamma }}{{d{F^2}}}\frac{\partial \varepsilon}{{\partial \gamma }},
\end{equation}
we obtain
\begin{equation}\label{epsilond}
\frac{{{\partial ^2}\varepsilon}}{{\partial {F^2}}} =\frac{{\sqrt {d - 1} }}{{2d{{\left[ {F\left( {1 - F} \right)} \right]}^{\frac{3}{2}}}}}\left[ {\ln \frac{{\gamma \left( {d - 1} \right)}}{{1 - \gamma }} - \frac{{2d\sqrt {F\left( {1 - F} \right)} }}{{\sqrt {d - 1} }}} \right].
\end{equation}

Now we only need to examine the sign of the term in the square bracket of the above equation, which is the same as that of $\frac{{{\partial ^2}\varepsilon}}{{\partial {F^2}}}$. In a change of variable, we let $x = \sqrt {\frac{{1 - F}}{{F\left( {d - 1} \right)}}} $, so that $x\in \left[ {0,1} \right]$ decreases monotonically with respect to $F$ in the interval $F \in \left[ {\frac{1}{d},1} \right]$. Let the term in the square bracket of Eq. \eqref{epsilond} be a function $f\left( x \right)$ of $x$, we have
\begin{equation}f\left( x \right) = \ln \left( {\frac{{dx}}{{1 - x}} + 1} \right) - \frac{{dx}}{{1 + \left( {d - 1} \right){x^2}}},\end{equation}
whose behavior at the two ends of the interval $x\in \left[ {0,1} \right]$ is easily found to be
\begin{align}&F \to \frac{1}{d},x \to 1,\frac{{{\partial ^2}\varepsilon}}{{\partial {F^2}}} \to  + \infty ;\label{limit1}\\
&F \to 1,x \to 0,\frac{{{\partial ^2}\varepsilon}}{{\partial {F^2}}} \to  - \infty .\label{limit2}\end{align}
Differentiate $f\left( x \right)$ with respect to $x$, we obtain
\begin{equation}
f'\left( x \right) = \frac{{\left( {d - 1} \right)\left( {d - 2} \right)dxg\left( x \right)}}{{\left( {1 - x} \right)\left( {dx - x + 1} \right){{\left( {1 + \left( {d - 1} \right){x^2}} \right)}^2}}},
\end{equation}
whose denominator is non-negative, and the function $g\left( x \right)$ is equal to
\begin{equation}
 g\left( x \right) = {{{\left( {x + \frac{2}{{d - 2}}} \right)}^2} - \frac{{{d^2}}}{{\left( {d - 1} \right){{\left( {d - 2} \right)}^2}}}} .\label{quadratic}\end{equation}
Noting that $g\left( x \right)$ is a quadratic function of $x$, it has two zeros which can be easily found to be
\begin{equation}{x_ \pm } = \frac{{ - 2\sqrt {d - 1}  \pm d}}{{\left( {d - 2} \right)\sqrt {d - 1} }}.\end{equation}
The zero ${x_ - } < 0$ is not in the interval $x\in \left[ {0,1} \right]$ of our interest, while it can be shown that ${x_ + }\in \left( {0,1} \right)$ if $d>2$. Furthermore, by evaluating the value of $g\left( x \right)$ in Eq. \eqref{quadratic} to determine the sign of $f'\left( x \right)$, we have
\begin{equation}
f'\left( x \right) \begin{cases}
 < 0&x \in \left( {0,{x_ + }} \right)\\
 \ge 0&x \in \left[ {{x_ + },1} \right)
\end{cases}.
\end{equation}

Thus we conclude that the minimum of the function $f \left( x\right)$ on the interval $x \in \left( {0,1} \right)$ is $f\left( x_ +\right)$
\begin{equation}
f\left( {{x_ + }} \right) =\ln \sqrt {d - 1} - \frac{{d - 2}}{{2\sqrt {d - 1} }} =h\left( {\sqrt {d - 1} } \right),
\end{equation}
where the function $h \left( x \right)$ is defined as
\begin{equation}h\left( x \right) \equiv \ln x - \frac{x}{2} + \frac{1}{{2x}}.\end{equation}
For $d>2$, we have $h'\left( x \right) =- \frac{{{{\left( {x - 1} \right)}^2}}}{{2{x^2}}} < 0$, and it follows that
\begin{equation}f\left( {{x_ + }} \right) < 0.\end{equation}
Combining this with the special case $f\left( 0 \right) = 0$, we have $f\left( x \right)<0$ and $ \frac{{{\partial ^2}\varepsilon}}{{\partial {F^2}}}  < 0$ for $x \in \left( {0,{x_ + }} \right)$. On the other hand, the function $f \left ( x \right)$ increases monotonically for $x \in \left[ {{x_ + },1} \right)$, and it can be easily shown that $f\left( 1 \right) \to  + \infty $ as $x \to 1$, which ensures that the function $f \left ( x \right)$ has only one zero in the interval $x \in \left( {0,1} \right)$. Also taking into account Eqs. \eqref{limit1} and \eqref{limit2}, which gives the limit of $ \frac{{{\partial ^2}\varepsilon}}{{\partial {F^2}}} $ as $F \to \frac{1}{d}$ and $1$, we conclude that the second derivative $ \frac{{{\partial ^2}\varepsilon}}{{\partial {F^2}}} $ has only one zero on the interval $F \in \left[ {\frac{1}{d},1} \right]$.

Now that we have proved that the function $\varepsilon \left( {F,d} \right)$ on the interval $F \in \left[{1,0}\right]$ is first concave upwards and then concave downwards as $F$ increases, the next step is simply to find a line both tangent to $\varepsilon \left( {F,d} \right)$ and passes the point $F=1$; in other words, we want to find a solution ${F_0} \in \left( {\frac{1}{d},1} \right)$ to the following equation
\begin{equation}\label{tipping}\log d - \varepsilon  = \left( {1 - F} \right)\frac{\partial \varepsilon}{{\partial F}} .\end{equation}
Taking the convexity of the function $\varepsilon \left( {F,d} \right)$ into account, one can deduce that such solution does exist and is unique. In addition, direct calculation shows that the conjecture $F=\frac{{4\left( {d - 1} \right)}}{{{d^2}}}$ is indeed the solution to Eq. \eqref{tipping}. Since $\varepsilon \left( {F,d} \right)$ can be easily shown to be monotonously increasing with respect to $F$, we now have adequate information to determine explicitly the value of $\mathcal{E}\left( {\rho _{\scriptscriptstyle{F}}^{iso}} \right)=\lim_{\alpha\to 1}{\mathcal{R}_{\alpha } }\left( {\rho _{\scriptscriptstyle{F}} ^{ iso }} \right)$,
\begin{equation}
\mathcal{E}\left( {\rho _{\scriptscriptstyle{F}}^{iso}} \right)=\label{T1app}
\begin{cases} 0&F \in \left[ {0,\frac{1}{d}} \right]\\
\varepsilon \left( {F,d} \right)&F \in \left[ {\frac{1}{d},\frac{{4\left( {d - 1} \right)}}{{{d^2}}}} \right]\\
\frac{{d\log \left( {d - 1} \right)}}{{d - 2}}\left( {F - 1} \right) + \log d&F \in \left[ {\frac{{4\left( {d - 1} \right)}}{{{d^2}}},1} \right]\end{cases}.
\end{equation}


\begin{thebibliography}{99}
\bibitem{01}M. A. Nielsen and I. L. Chuang, \textit{Quantum Computation and Quantum Information} (Cambridge University
Press, New York, 2000).

\bibitem{mixedstate1}C. H. Bennett, G. Brassard, S. Popescu, B. Schumacher,
J. Smolin, and W. K. Wootters, Phys. Rev. Lett
78, 2031 (1996),

\bibitem{mixedstate2}C. H. Bennett, D. P. DiVincenzo, J. A. Smolin, and W. K. Wootters,
Phys. Rev. A {\bf 54}, 3824 (1996).


\bibitem{VedralPlenio}V. Vedral, M. B. Plenio, M. A. Rippin, and P. L. Knight,
Phys. Rev. Lett. {\bf 78}, 2275 (1997).




\bibitem{VedralRMP}V. Vedral,
Rev. Mod. Phys. {\bf 74}, 197 (2002).

\bibitem{PlenioVirmani}M. B. Plenio and S. Virmani,
Quant. Inf. Comput. 7, 1 (2007).


\bibitem{HorodeckiE}R. Horodecki, P. Horodecki, M. Horodecki, and K. Horodecki,
Rev. Mod. Phys. {\bf 81}, 865 (2009).









\bibitem{NielsenEM}M. A. Nielsen,
Phys. Rev. A {\bf 61}, 064301 (2000).

\bibitem{JPLOCC}D. Jonathan and M. B. Plenio,
Phys. Rev. Lett. {\bf 83}, 1455 (1999).

\bibitem{Vidalmixedqubits}G. Vidal,
Phys. Rev. A {\bf 62}, 062315 (2000).

\bibitem{VidalEM}G. Vidal,
J. Mod. Opt. {\bf 47}, 355 (2000).


\bibitem{Turgut}S. Turgut, J. Phys. A: Math. Theor. {\bf 40}, 12185 (2007)

\bibitem{Klimesh}M. Klimesh,  arXiv.0709.3680 (2007).




\bibitem{Cui1}J. Cui, M. Gu, L. C. Kwek, M. F. Santos, H. Fan and V. Vedral,
Nature Commun. {\bf 3}, 812 (2012).

\bibitem{Cui2}J. Cui, L. Amico, H. Fan, M. Gu, A. Hamma, and V. Vedral,
Phys. Rev. B {\bf 88}, 125117 (2013).


\bibitem{Cui3}F. Franchini, J. Cui, L. Amico, H. Fan, M. Gu, L. C. Kwek,
V. Korepin and V. Vedral,
Phys. Rev. X {\bf 4}, 041028 (2014).

\bibitem{Alioscia1}S. T. Flammia, A. Hamma, T. L. Hughes, and X. G. Wen,
Phys. Rev. Lett. {\bf 103}, 261601 (2009).

\bibitem{Alioscia2}G. B. Halasz and A. Hamma, Phys. Rev. Lett. {\bf 110}, 170605 (2013).

\bibitem{Alioscia3}A. Hamma, L. Cincio, S. Santra, P. Zanardi, and A. Luigi,
Phys. Rev. Lett. {\bf 110}, 210602 (2013).




\bibitem{WoottersEoF}W. K. Wootters,
Phys. Rev. Lett {\bf 80}, 2245 (1998).


\bibitem{KSRenyi}J. S. Kim and B. C. Sanders,
J. Phys. A {\bf 43}, 445305 (2010).


\bibitem{CKWmonogamy}V. Coffman, J. Kundu, and W. K. Wootters,
Phys. Rev. A {\bf 61}, 052306 (2000).

\bibitem{OVmonogamy}T. J. Osborne and F. Verstraete,
Phys. Rev. Lett. {\bf 96}, 220503 (2006).


\bibitem{Wernerstate}R. F. Werner,
Phys. Rev. A {\bf 40}, 4277 (1989).

\bibitem{TerhalEoF}B. M. Terhal and K. G. H. Vollbrecht,
Phys. Rev. Lett. {\bf 85}, 2625 (2000).

\bibitem{RCconcurrence}P. Rungta and C. M. Caves,
Phys. Rev. A {\bf 67}, 012307 (2003).

\bibitem{VWEM}K. G. H. Vollbrecht and R. F. Werner,
Phys. Rev. A {\bf 64}, 062307 (2001).



\bibitem{HWconcurrence}S. Hill and W. K. Wootters,
Phys. Rev. Lett. {\bf 78}, 5022 (1997).






\end{thebibliography}
\end{document}